# A Monolithic Graphene-Functionalized Microlaser for Multispecies Gas Detection

*Yanhong Guo, Zhaoyu Li, Ning An, Yongzheng Guo, Yuchen Wang, Yusen Yuan, Hao Zhang, Teng Tan, Caihao Wu, Bo Peng, Giancarlo Soavi,\* Yunjiang Rao,\* and Baicheng Yao\**

Optical-microcavity-enhanced light–matter interaction offers a powerful tool to develop fast and precise sensing techniques, spurring applications in the detection of biochemical targets ranging from cells, nanoparticles, and large molecules. However, the intrinsic inertness of such pristine microresonators limits their spread in new fields such as gas detection. Here, a functionalized microlaser sensor is realized by depositing graphene in an erbium-doped over-modal microsphere. By using a 980 nm pump, multiple laser lines excited in different mode families of the microresonator are co-generated in a single device. The interference between these splitting mode lasers produce beat notes in the electrical domain (0.2–1.1 MHz) with sub-kHz accuracy, thanks to the graphene-induced intracavity backward scattering. This allows for lab-free multispecies gas identification from a mixture, and ultrasensitive gas detection down to individual molecule.

## 1. Introduction

In the past two decades, whispering-gallery-mode (WGM) optical microresonators have attracted enormous interests, owing to their unique high quality factors ($Q$) and small mode volumes.[1–3] The enhanced light–matter interactions in a WGM microcavity enable the development of microlasers, microsensors, and microspectrometers.[4–9] In particular, WGM microcavities offer unique advantages when applied to the fields of biological and chemical sensing, where they provide ultrahigh frequency resolution in sub-MHz level[10] by measuring the passive interference between different cavity modes[11] as well as the shift and/or broadening of the mode resonances.[12] With such approach, detection of individual cells,[13,14] viruses,[15] protein molecules,[16] DNA,[17,18] and nanoparticles[19,20] has already been achieved. Further enhancement of the performances of a WGM-based sensor can be obtained in an active cavity, where the monolithic integration of a laser provides higher stability and improved frequency resolution.[21,22] Such platforms hold great promise also for applications in the field of gas sensing, however, the materials used in conventional WGM microcavities (silica or metal fluorides) are inert and thus unsuitable for gas adsorption and tracing. On the other hand, the hybridization of 2D materials with microcavities, such as chip microrings,[23–25] plasmonic resonators,[26,27] nanowires,[28] fiber microcavities,[29,30] and WGM microresonators,[31–33] offers a new route for enhanced photon–electron interactions and thus provides a novel platform for gas detection.[34]

Here we realize an active WGM microsphere laser device functionalized with single-layer graphene that is capable of

Y. Guo, Z. Li, N. An, Y. Wang, Y. Yuan, H. Zhang, T. Tan, C. Wu, Y. Rao, B. Yao
Key Laboratory of Optical Fibre Sensing and Communications
(Education Ministry of China)
University of Electronic Science and Technology of China
Chengdu 611731, China
E-mail: yjrao@uestc.edu.cn; yaobaicheng@uestc.edu.cn
Y. Guo, B. Peng
State Key Laboratory of Electronic Thin Film and Integrated Devices
University of Electronic Science and Technology of China
Chengdu 611731, China

The ORCID identification number(s) for the author(s) of this article can be found under https://doi.org/10.1002/adma.202207777.

Y. Yuan
School of Engineering
University of Glasgow
Glasgow G12 8QQ, UK
G. Soavi
Institute of Solid State Physics
Friedrich Schiller University Jena
07743 Jena, Germany
E-mail: giancarlo.soavi@uni-jena.de
G. Soavi
Abbe Center of Photonics
Friedrich Schiller University Jena
07745 Jena, Germany
Y. Rao
Research Centre for Optical Fiber Sensing
Zhejiang Laboratory
Hangzhou 310000, China

**DOI: 10.1002/adma.202207777**

multispecies gas detection (i.e., the capability to simultaneously distinguish $CO_2$, $NH_3$, $NO_2$, and $H_2O$ molecules in a mixture) with high sensitivity (i.e., individual molecule). Thanks to the over-modal nature of our device, we successfully generate multiple laser modes in one single device. By depositing graphene 20° away from the equator, we make it interact with the higher order modes of the laser microcavity, while avoiding thermal damage. The backward scattering induced by the graphene monolayer induces symmetry breaking and mode splitting[35] at specific frequencies, which can be precisely measured via optoelectronic heterodyne beating. In addition, gas molecule adsorption on graphene changes its Fermi level,[36,37] and thus modifies the effective permittivity of the cavity[38] leading to tens of kHz/ppb changes in the mode splitting. From this, we can precisely measure the presence of various gases in a mixture with high precision, discrimination, and sensitivity.

## 2. Results

**Figure 1**a shows the conceptual design of our graphene-based microlaser sensor (GMLS). A silica microsphere with diameter of ≈600 μm is doped with erbium by using the solution coating and the discharge sintering technology,[39] leading to an effective erbium concentration $>10^{18}$ cm$^{-3}$. Subsequently, we deposit a mechanically exfoliated graphene flake by deterministic

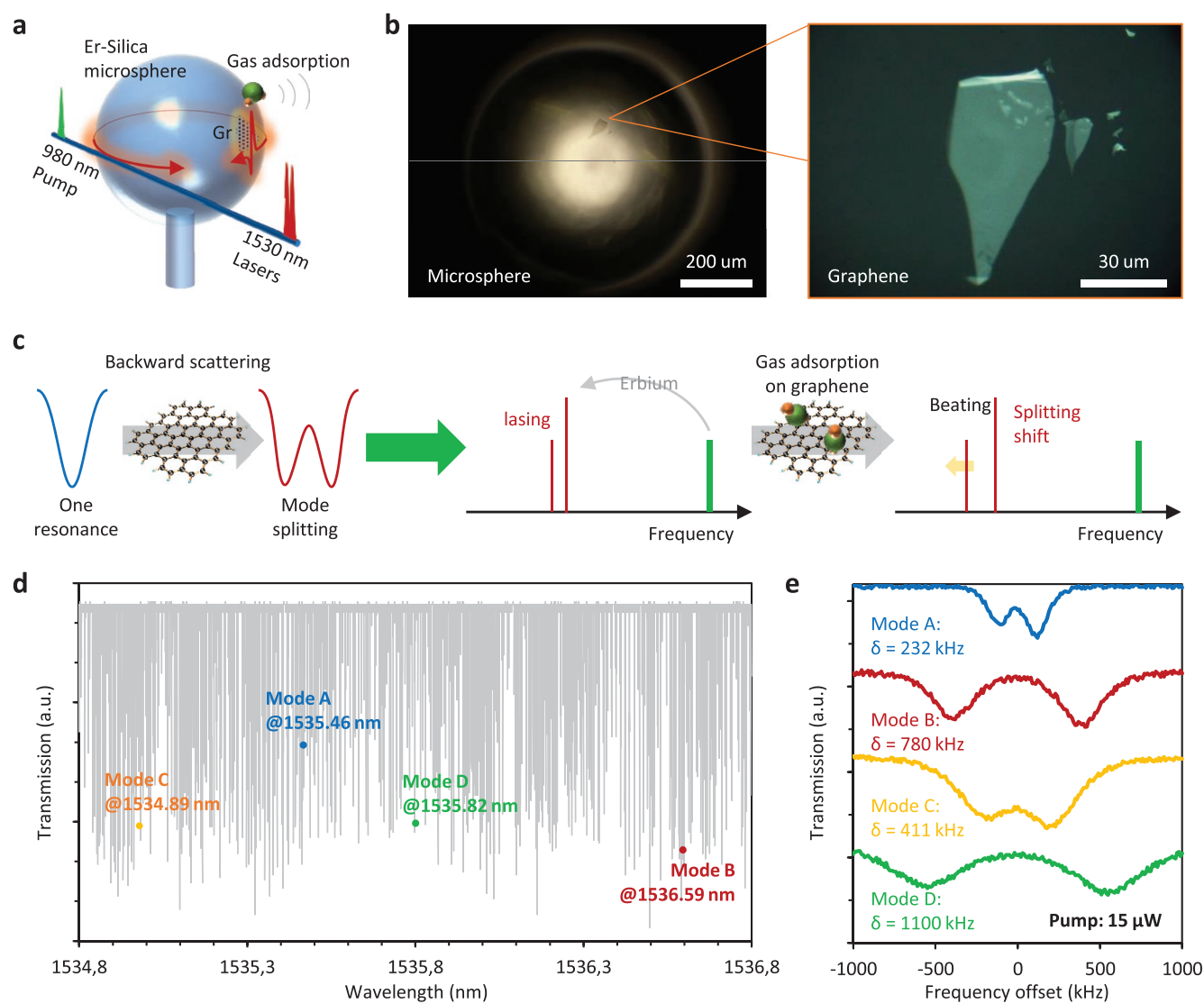

**Figure 1.** Conceptual design of the graphene-based active microresonator and the laser-splitting-based gas detection. a) Schematic diagram of the device. A graphene layer is deposited on the WGM microsphere. A 980 nm diode laser (ECDL) is used as the pump to excite the erbium microcavity lasers, which shows mode slitting due to the graphene back scattering. The splitting frequencies are sensitive to gas–graphene interactions. b) Optical microscopy images show the exfoliated graphene deposited on the microsphere, with area ≈30 × 60 μm². c) Process of the optical response. First, graphene induces mode splitting, then two laser lines are generated in the splitting mode, finally the gas adsorbed on graphene changes the splitting offset. d) Measured transmission spectrum of the graphene-based active microresonator. Here we mark four selected spitting modes due to the graphene back scattering. e) Details of the four selected modes A–D. In order to observe the splitting, we tune the pump power below the lasing threshold for loss compensation.

dry-transfer at 20° above the equatorial plane, to ensure the light–material interaction with high order modes while avoiding heating damage. Next, when we launch a 980 nm pump laser into the microcavity via a tapered fiber, we obtain a counter-propagating laser with splitting frequencies. In this configuration, gas adsorption on the graphene will change the permittivity of the device and thus modify the splitting offsets. More details about the operating principle of our device are discussed in Note S1, Supporting Information. Figure 1b shows the side-view microscopic picture of our device and a zoomed-in image where the $30 \times 60\,\mu m^2$ graphene layer is clearly visible. In comparison to erbium-doped silica (i.e., the material of the microcavity), graphene has a higher refractive index and its dimensions on the sphere are much larger compared to the optical wavelength, thus enabling strong backward scattering for the mode splitting formation. More details about the device nanofabrication, graphene transfer, and material characterization are available in the Note S2, Supporting Information. In Figure 1c, we schematically depict the sensing mechanism of our device. First, due to the presence of graphene and its induced backward scattering, some modes of the laser cavity will split, forming a pair of symmetric and asymmetric standing waves with splitting offset $\delta$. Then, the adsorption of gas molecules on graphene will change its permittivity and thus shift the mode splitting from $\delta$ to $\delta'$, which can be eventually precisely measured by heterodyne detection in the radio-frequency domain.

Figure 1d plots the passive transmission spectrum of our graphene-based microresonator. In one single free-spectral-range (100 GHz), there are 517 resonances. Before graphene deposition, the typical $Q$ factor of the erbium-doped microsphere at its fundamental mode ($TM_{01}$) is $\approx 8 \times 10^7$. After graphene deposition, the average $Q$ factor decreases by ≈12%, which is still enough for lasing with approximately tens of µW pump power. By sending the pump power for loss compensation but still remaining below the laser threshold (e.g., 15 µW), we can measure the cavity transmission with narrower resonances and thus get more insights into the mode splittings.[22,40] We find that several splitting modes appear after the deposition of graphene, such as modes A–D, located at 1535.46, 1536.59, 1534.89, and 1535.82 nm. Figure 1e shows the four selected splitting modes (A–D), with splitting offset $\delta$ of 232, 780, 411, and 1100 kHz, respectively. We note that the splitting offsets of our device are smaller than many WGM microcavities from previous publications,[10,22] mainly because the geometric size (or mode volume) of our microsphere is larger. The observed kHz splitting offset is also an evidence that the double resonances are caused by mode splitting rather than mode crossing. In Note S3, Supporting Information, we provide more details about the splitting resonances and their temporal traces. We also note that mode splitting in this microsphere resonator can be induced by other scatterers (such as inhomogeneous erbium doping or scattering points inside the spherical cavity). We can observe these splitting modes before the graphene deposition, but they are insensitive to gas molecules (see Notes S2 and S3, Supporting Information).

**Figure 2**a shows the lasing behavior of our GMLS device. By increasing the 980 nm pump laser power, different laser lines are gradually excited in the communication band around 1535 nm. The lasing threshold of this device (after deposition of graphene) is 16 µW, and the lasing efficiency is ≈1%. While increasing the pump power, we track in particular four states (i–iv in Figure 2), which correspond to the formation of the laser modes A–D. Figure 2b plots the optical spectra of these four states and Figure 2c plots the radio frequency spectra (i.e., the splitting offset) of the four states: the modes A–D (states i–iv) have central wavelengths of 1535.462, 1536.614, 1534.913, and 1535.826 nm and they have splitting offsets of 231.8, 780.4, 411.5, and 1100 kHz, respectively. All these laser modes are in agreement with the transmission measurement, as the intra-cavity laser induced heating is very low, and we use a thermo-electrical-cooler (TEC) for temperature stabilization. We also note that additional laser modes can be obtained by tuning the pump polarization (see Note S3, Supporting Information). In the following we will focus only on modes A–D.

Figure 2d shows a further characterization of the four beat notes for a pump power of 40 µW. For the laser modes A–D, we observe signal to noise ratios (SNR) of 43, 31, 18, and 23 dB, and linewidths of 76, 107, 115, and 183 Hz, respectively. These values of the SNR and linewidths allow to achieve high energy and frequency resolution for gas sensing applications far beyond the typical performances of passive devices. We also measure the spectral uncertainties of the four modes, and plot them in Figure 2e. For measurements up to 5 min, the maximum drifting of modes A–D is ±510, ±645, ±390, and ±870 Hz, respectively. Thus, for the gas sensing application that we discuss in the following, we will refer to the largest measured instability of each mode for an estimate of the detection limit.

Next, we demonstrate the performances of our device for gas sensing applications (**Figure 3**). First, we schematically reiterate the sensing mechanism in Figure 3a: the out-of-plane $\pi$ bonds in graphene are highly sensitive to polar gases.[41,42] When adsorbed on graphene, gases such as $NO_2$ and $H_2O$ act as electron acceptors, whereas gases such as $NH_3$ and $CO_2$ act as electron donors.[36,43] Thus, starting from our originally p-doped graphene on silica ($E_F \approx 0.2$ eV, as obtained from Raman spectroscopy shown in Note S2, Supporting Information),[45] $NO_2$ and $H_2O$ adsorption will increase the $|E_F|$, while vice versa the $NH_3$ and $CO_2$ adsorption will decrease the $|E_F|$. In the Supporting Information, we also confirm the electrical response of a graphene device when exposed to such gases.[44]

Figure 3b shows the calculated permittivity of graphene, as a function of both $|E_F|$ and the optical wavelength. In our experiment, we start from $|E_F| \approx 0.2$ eV and wavelength ≈1535 nm. Changes of the graphene's permittivity directly affect the mode splitting, following Equation (1)[35]

$$\delta = \frac{-\varepsilon_m \, \mathrm{Re}[\Psi] f^2(r) \omega_c}{\varepsilon_c V_m} \tag{1}$$

where $\Psi = 3V_s(\varepsilon_g - \varepsilon_m)/(\varepsilon_g + 2\varepsilon_m)$ is the scattering factor induced by the graphene, $f^2(r)$ is field spatial variation defined by the optical wavelength, $\varepsilon_m$, $\varepsilon_g$, and $\varepsilon_c$ are the ambient permittivity, the graphene permittivity, and the cavity permittivity, respectively. $V_s$ is the effective size of the graphene scatterer ($30 \times 60\,\mu m^2$), $\omega_c$ is the resonant frequency of the cavity, and $V_m$ is the cavity mode volume (more details are given in Note S1,

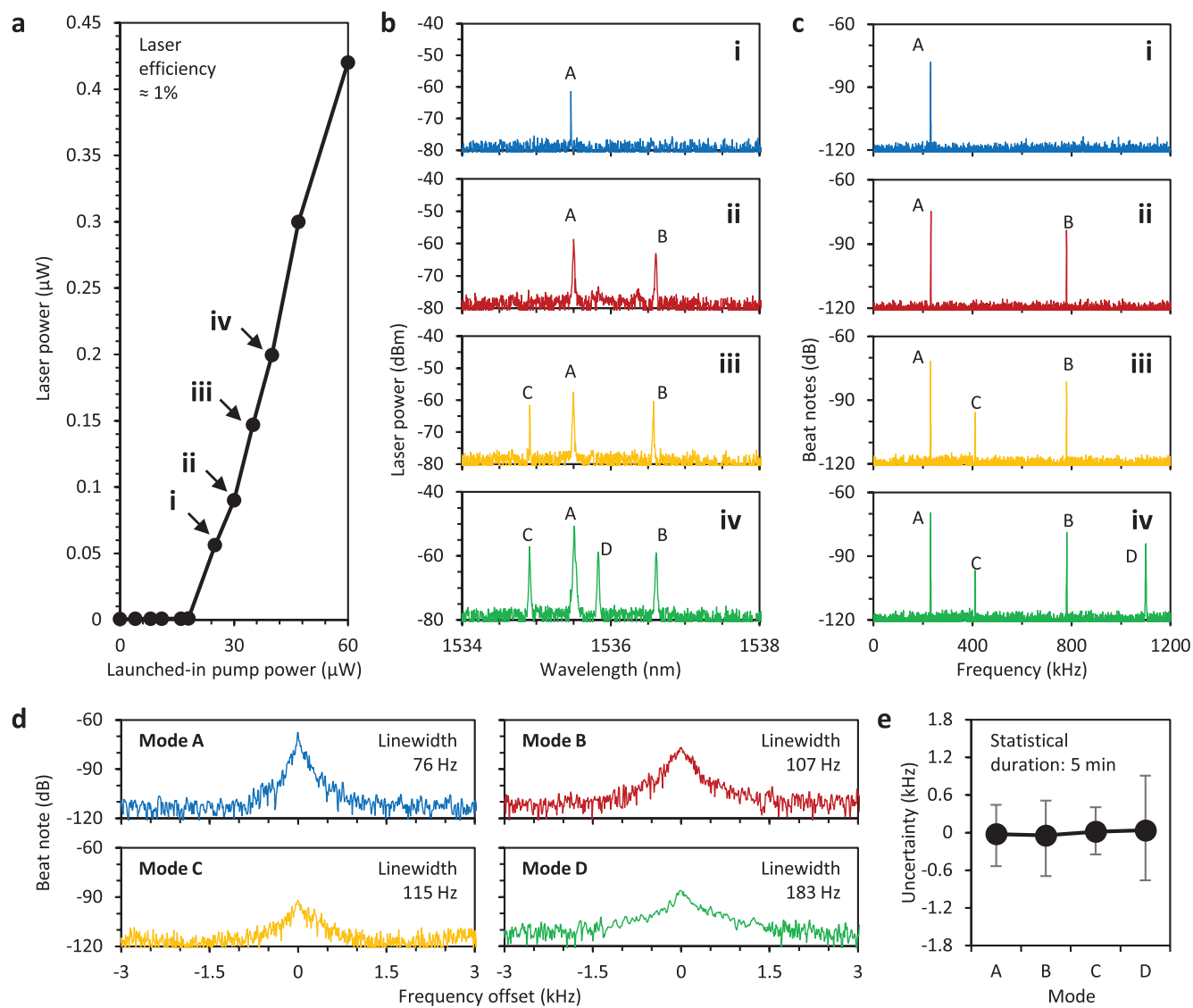

**Figure 2.** Laser operation of the device. a) Correlation between the pump power and the lasing power, measured by using an optical spectral analyzer. The lasing threshold is 16 µW, and the lasing efficiency ≈1%. b,c) Optical and electrical spectra of the laser lines. By increasing the pump power, multiple laser modes are observed. Here we focus on the four modes A–D. When the pump power reaches 40 µW, the laser modes A–D are co-excited in the 1535 nm band. Each selected mode has two splitting branches, with self-beat notes in the RF region 0~1200 kHz. d) Zoomed in spectra of the four selected beat notes from (c). They all show high SNR and approximately hundreds Hz linewidth, due to their laser nature. e) Spectral uncertainty of the modes A–D, measured over a 5 min time window.

Supporting Information). As a consequence, the increment (decrement) of $\varepsilon_g$ enables an increase (decrease) of the beating frequency of a splitting mode. Figure 3c shows a sketch of our experimental setup. A 980 nm laser diode drives the GMLS. In front of the device (gas sensor), a fiber polarization controller is used to optimize the mode excitation and an isolator is used to avoid back-scattering of light inside the pump laser. The GMLS is fixed in a gas chamber with in–out gas channels and a TEC is connected to the microcavity for thermal stabilization. We use a tapered fiber to inject and collect optical signals. In order to optimize the effect of pumping, the microcavity works in the over-coupling region. Then we use a WDM to filter out the laser lines and thus to obtain higher SNR. Finally, the laser lines are measured by using an optical spectrum analyzer (OSA) and an electrical spectrum analyzer (ESA). For the detection of low concentrations of gases in a mixture, we mainly trace the spectral shifts of the laser beat notes in the ESA.

The measured beating frequency shifts induced by $CO_2$, $NH_3$, $NO_2$, and $H_2O$ in modes A–D as a function of gas concentration are shown in Figure 3d (see also the spectra in Note S3, Supporting Information). For instance, when increasing the $CO_2$ concentration from 0 to 1000 ppb, the frequency of the beat notes of the four selected modes (A–D) decreases by 11, 20.4, 16, and 26.8 kHz; while when increasing the $NO_2$ concentration from 0 to 1000 ppb, the beat notes increase by 14, 33, 18, and 63 kHz. Considering the abovementioned frequency uncertainties of the four selected beat notes, we can estimate the detection limit of the four gases in our device to be 0.5, 0.8, 0.3 and 20 ppb for $CO_2$, $NH_3$, $NO_2$, and $H_2O$, respectively, as shown in Figure 3e. The lower sensitivity to $H_2O$ compared to the other gases may be due to the loose molecular binding of $H_2O$ and graphene.[36] The different sensitivity coefficients, obtained from a logarithmic fit of the curves in Figure 3d, allow to simultaneously detect different beating frequencies, and thus to identify different gases in a mixture. Thus, in order to identify the four investigated gases in a mixture with high selectivity, we simply need to solve the system of linear Equation (2):

$$\begin{bmatrix} \Delta f_A \\ \Delta f_B \\ \Delta f_C \\ \Delta f_D \end{bmatrix} = \begin{bmatrix} -2.65 & -4.39 & 3.39 & 1.08 \\ -5.54 & -7.37 & 7.98 & 2.49 \\ -4.01 & -6.33 & 4.07 & 1.57 \\ -7.70 & -10.55 & 15.99 & 4.12 \end{bmatrix} \begin{bmatrix} \log_{10}(C_{CO_2}) \\ \log_{10}(C_{NH_3}) \\ \log_{10}(C_{NO_2}) \\ \log_{10}(C_{H_2O}) \end{bmatrix} \quad (2)$$

where $C_{CO_2}$, $C_{NH_3}$, $C_{NO_2}$, and $C_{H_2O}$ are the different gas concentrations.

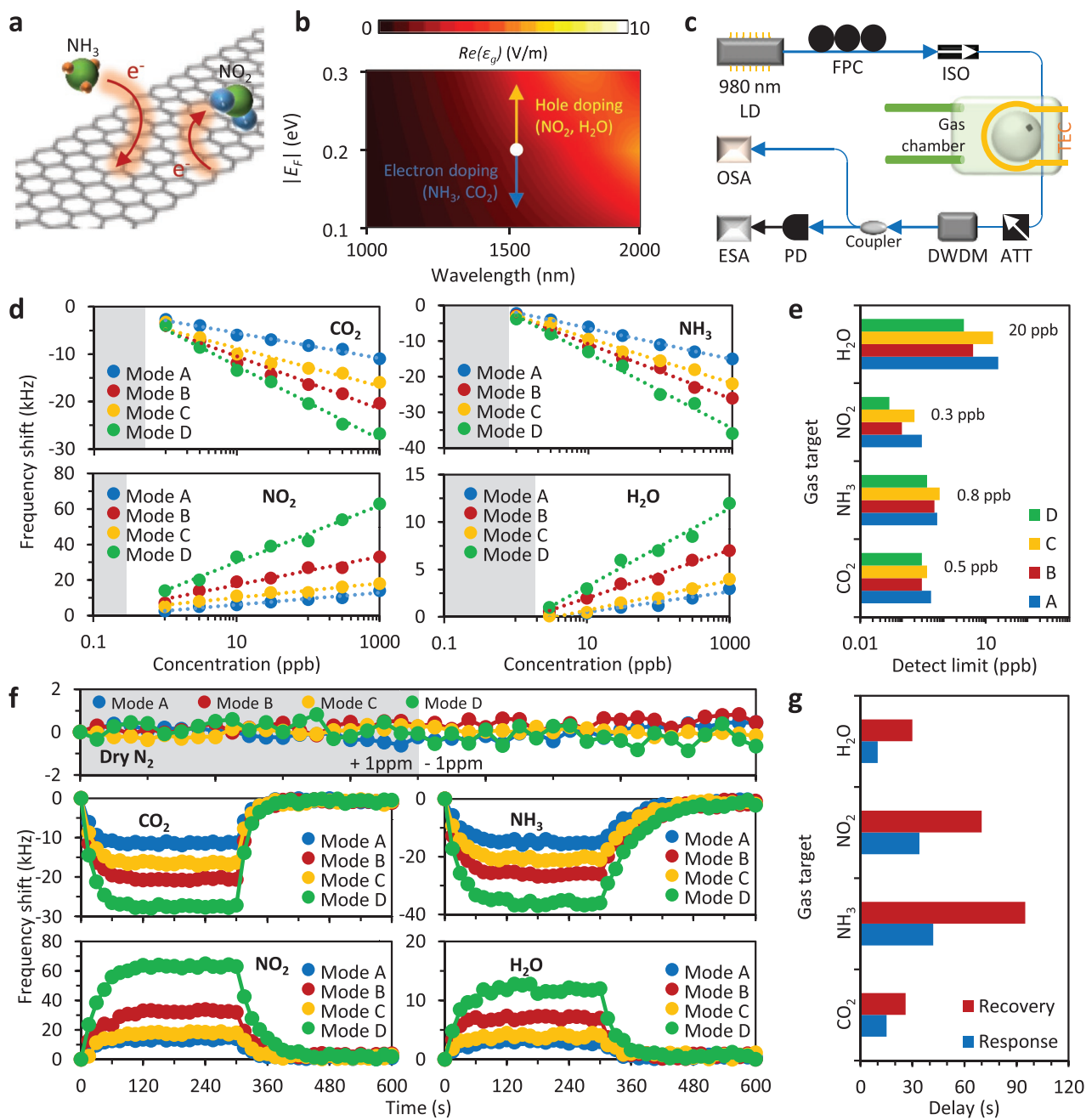

**Figure 3.** Optoelectronic detection of four gas samples. a) Sketch of the graphene–gas interaction. Polar gases adsorbed on graphene can act as electron donors/acceptors, thus tuning the Fermi level of graphene. b) Calculated permittivity of graphene ($\varepsilon_g$), which is defined by its Fermi level. c) Experimental setup for gas sensing. LD: laser diode, FPC: fiber polarization controller, ISO: isolator, TEC: thermal–electrical cooler, ATT: optical attenuator, WDM: wavelength division de-multiplexer, PD: photodetector, OSA: optical spectrum analyzer, ESA: electrical spectrum analyzer. d) Measured correlation of the gas concentration and the frequency shifts of the beat notes. The four panels show the cases when measuring $CO_2$, $NH_3$, $NO_2$, and $H_2O$ gases. The colored dots show the results of modes A–D. Different splitting beat notes show different "concentration–shift" correlations (sensitivity coefficients). e) Estimated detection limit extracted from the laser instabilities at the different laser lines. f) Measured spectral response and recovery for the four selected gases. The device recovery for all the gas samples is almost 100%. In contrast, this sensor is not sensitive to dry $N_2$. g) Response and recovery time. Specifically, the material limited response (recovery) time is less than 45 s (95 s) for $NH_3$.

In Figure 3f, we further demonstrate the recoverability of the GMLS. First we confirm that the mode responses are insensitive to dry $N_2$: when the concentration of dry $N_2$ changes by ±1 ppm, splitting frequencies keep unchanged. Then, by periodically injecting 1 ppm target gas and dry $N_2$, we monitor the frequency shifts of the beat notes. We use the gas syringe with a fixed injecting speed. Every 10 min period, the gas is injected in for 60 s, we keep the chamber stable for 240 s (temperature 300 K), and then slowly discharge the gas in 300–360 s by injecting dry $N_2$. Finally we keep the chamber stable again until 600 s. For each gas sample, our device demonstrates high recoverability (almost 100%). As shown in Figure 3g, when the gas concentration is stable, both the response delay and the recovery delay are in the single minute level. The response/recovery time of our device depends mainly on three factors: the gas concentration (*N*), the ambient temperature (*T*), and the bonding efficiency (*ξ*), which is defined by the gas kinetic theory.[46] The average frequency of gas adsorption on the graphene surface is $F = A\xi(N/4)(8k_B T/\pi m)^{1/2}$ (unit: molecules/s), where *A* is the area of the graphene, *N* is the gas concentration, $k_B$ is the Boltzmann's constant, and *m* is the gas molecular mass.[47] The *ξ* would decrease to 0 when the gas adsorption approaches saturation. We thus expect that a higher injected gas concentration (*N*) or a higher temperature (*T*) can both contribute to accelerate the response/recovery time (see also Note S3, Supporting Information, for details).

Moreover, this scheme using multiple laser modes allows us to quantitatively measure the concentrations of the four gases in the mixture via solving the linear equations. As a proof of concept, we use the device to detect four different gas mixtures (consisting of $CO_2$, $NH_3$, $NO_2$, and $H_2O$) with component ratios: (i) 1:0.1:0.1:0.1, (ii) 0.1:1:0.1:0.1, (iii) 0.1:0.1:1:0.1, and (iv) 0.1:0.1:0.1:1 in ppm scale (gray columns) and we measured: (i) 0.93:0.105:0.11:0.12, (ii) 0.088:1.11:0.109:0.093, (iii) 0.085:0.113:0.92:0.11, and (iv) 0.11:0.12:0.13:0.89 (red columns). The statistic error could be further reduced via repeated measurements (**Figure 4**a,b). Besides simultaneously measuring four gas samples, our GMLS is also capable to quantitatively detect mixtures with two or three gas species. For three gas mixtures, we flexibly mix $CO_2/NH_3/NO_2$, $CO_2/NH_3/H_2O$, $CO_2/NO_2/H_2O$, and $NH_3/NO_2/H_2O$, and we obtain accurate results (with detection error <7%) by detecting the shifts of three laser modes (e.g., B–D), as shown in Figure 4c,d. The same can be done for gas mixtures containing two different gases (Figure 4e,f): in this case only two laser modes are necessary (e.g., B and D) and the average error in the double-gas detection is <5%.

Moreover, the sub-ppb sensitivity of our device points toward the possibility of measuring individual molecule dynamics. The size of the transferred graphene flake ($\approx 1.5 \times 10^{-9}$ m$^2$) is much smaller than the gas chamber ($8 \times 10^{-3}$ m$^3$), and we thus estimate that there are <1000 gas molecules per second interacting with the graphene sheet when the gas concentration is <1 ppb. To trace the individual molecular dynamics, we build a lock-in amplification setup,[17,34] as shown in **Figure 5**a. Here, we select the laser mode D by using an optical filter (FWHM 0.1 nm). After a photodetection, the beat note of mode D is mixed with a stable electrical signal for further frequency down conversion. This step is needed to reach a frequency (<100 kHz) that can be detected by our lock-in amplifier. Finally, we send the mixed signal ($f_{mix} = 78$ kHz) to the lock-in amplifier and set the lock-in bandwidth to 100 Hz. Figure 5b plots the $f_{mix}$ signal before and after injecting 0.1 ppb of $NO_2$ gas into the chamber (originally filled with pure $N_2$). The gas adsorption induces a beating frequency shift, and thus the amplitude decreases at the frequency detected by the lock-in.

Such intensity changes are monitored in an oscilloscope (Figure 5c). In the 0–5 s (blue region), the sensor is exposed to pure $N_2$, while in the 5–15 s (orange region), the sensor is exposed to pure $N_2$ + 1 ppb $NO_2$. The intensity resolution in this measurement is ±8.33 µV. When the chamber is filled with $N_2$, the lock-in trace is flat. In contrast, with 1 ppb $NO_2$ the intensity of the lock-in signal is dynamically modulated around an average value of 0.8 mV, due to individual molecule $NO_2$–graphene interactions. In Figure 5d, we zoom-in these temporal dynamics and compare the two states i (pure $N_2$,) and ii (1 ppb $NO_2$). The minimum step height in state ii is ±50 µV, and the height of any other step is an integer multiple of this value, supporting the hypothesis that we are detecting individual molecule dynamics. We thus measured the distribution of 200 different steps in state ii and found that the statistics follows a power law (Figure 5e), as expected in the case of individual molecule adsorption and desorption events.[34,37]

## 3. Discussion

We have shown that our graphene microlaser sensor holds great potential for multispecies gas detection with compact size and low power consumption. However, before a transition from lab to fab, there are still few foreseeable challenges: 1) the gas detection relies on the specifications of each device, which should thus be calibrated before use; 2) this sensor can only quantitatively measure the identified (calibrated) gas mixtures. To address these issues, we suggest the following schemes: 1) characterize and fix a set of fabrication parameters (such as standardized graphene size/location and WGM cavity parameters); 2) fix the fiber–cavity coupling to optimize the nanofabrication as consistent as possible.[34] Moreover, each sensor could be pre-inspected by using artificial intelligence methods.[48] In the Note S3, Supporting Information, we demonstrate that we can fabricate several devices with controlled size and position of the exfoliated graphene flake. Moreover, we show a plug and play elementary prototype of our GMLS. In addition, for the detection of additional gases (more than only the four presented here), one could simply boost the laser power and excite more splitting laser modes. This will lead to a more complex demodulation scheme, which could be addressed by future and more advanced optical signal processing.[27]

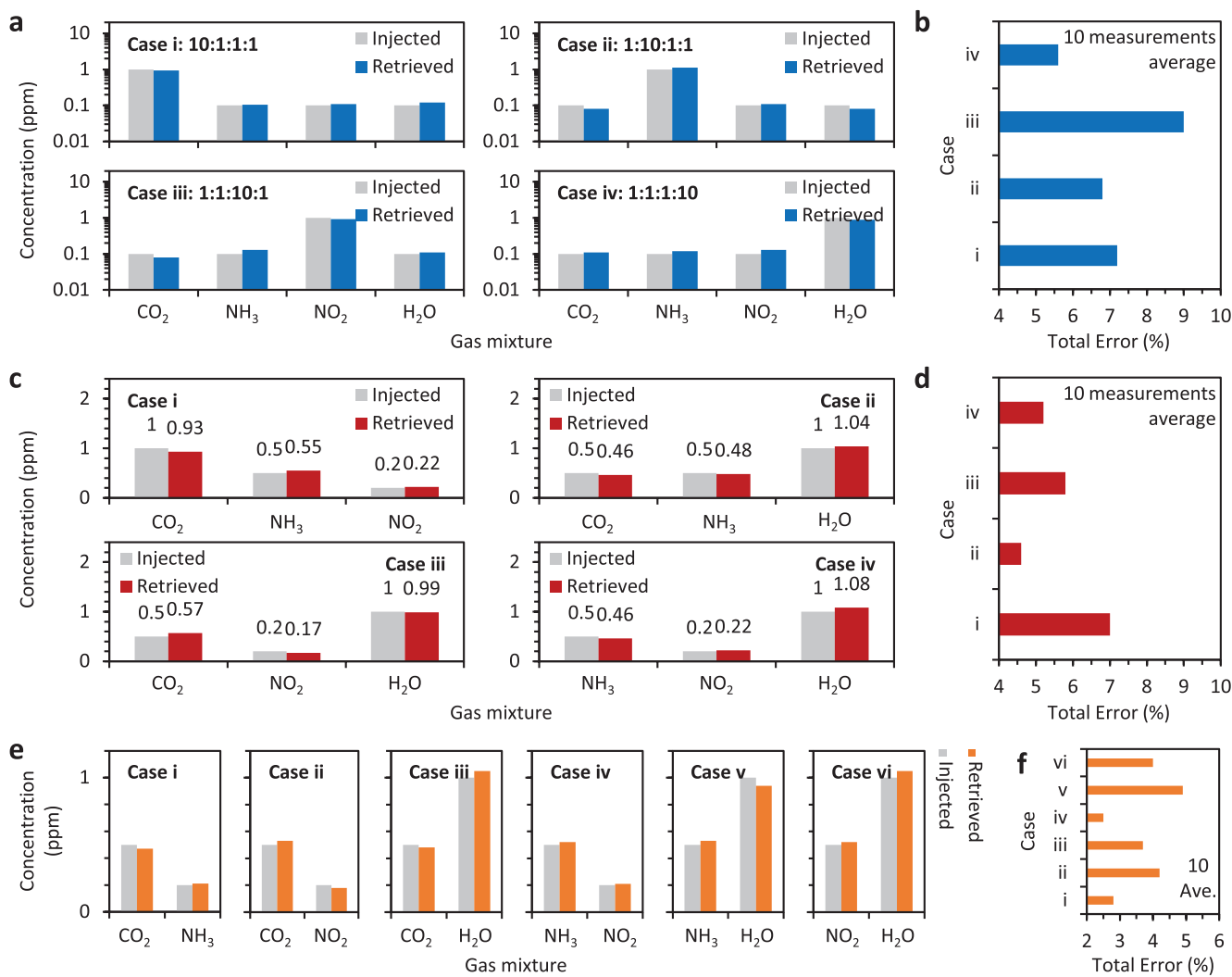

**Figure 4.** Detection of gases in mixtures. a) Detection of four selected gases in mixtures, by using modes A–D. b) Average errors of four gas detection over ten measurements. c) Detection of three selected gases in mixtures, by using modes B–D. d) Average errors of three gas detection over ten measurements. e) Detection of two selected gases in mixtures, by using mode B and D. f) Average errors of two gas detection over ten measurements. In (a), (c), and (e), the grey columns show the real proportions, and the colored columns show the results measured by our laser sensor. In (b), (d), and (f), the errors could be further optimized by increasing the number of repeated measurements.

To conclude, in this work we demonstrated the generation of multiple-splitting-mode lasers in a graphene-based active microresonator, offering an all-optical sensor for multispecies gas detection with sub-ppb sensitivity. By placing a single layer graphene 20° away from the equator of the microresonator, we found and measured laser splitting offsets that are extremely sensitive to adsorption of polar gases on graphene. The beating signals of such frequency offsets can be conveniently measured in the RF spectrum. By leveraging the narrow and stable linewidths of our laser device, we achieved sub-kHz spectral resolution and the consequent gas identification at the ppb level concentration. Moreover, the presence of multiple beats in one single device allows for the identification of different gases in a mixture. This scheme offers a label-free optical tool to realize both qualitative and quantitative gas molecule detection, with compact size, low power consumption, and simple operation.

## 4. Experimental Section

*Theoretical Analysis*: In a whispering-gallery-mode microcavity, each travelling mode contained a pair of counter-propagating waves with degenerate frequency. An intracavity back scattering could break this degeneracy and form a mode splitting, described by the Heisenberg equation.[35] Such mode splitting depended on the permittivity of the scatterer (graphene in this work). In Note S1, Supporting Information, the physics behind the mode splitting, its effect on the laser, and the mechanism of the graphene-based optical gas sensing are discussed.

*Nanofabrication of the Graphene/Erbium–Silica Microresonator*: The active microsphere samples were fabricated with the arc-discharge technique in a fiber fusion splicer (FITEL S178), and the erbium ions were introduced via the solution coating and the discharge sintering technology. By optimizing the fabrication parameters, uniform microspheres with diameter of 600 μm and surface erbium doping rate $10^{18}$ cm$^{-3}$ were obtained. High-quality crystalline single layer graphene nanosheets were dry transferred onto the microsphere samples with the PDMS stamping method. In Note S2, Supporting Information, details of the nanofabrication process and the related characterization are shown.

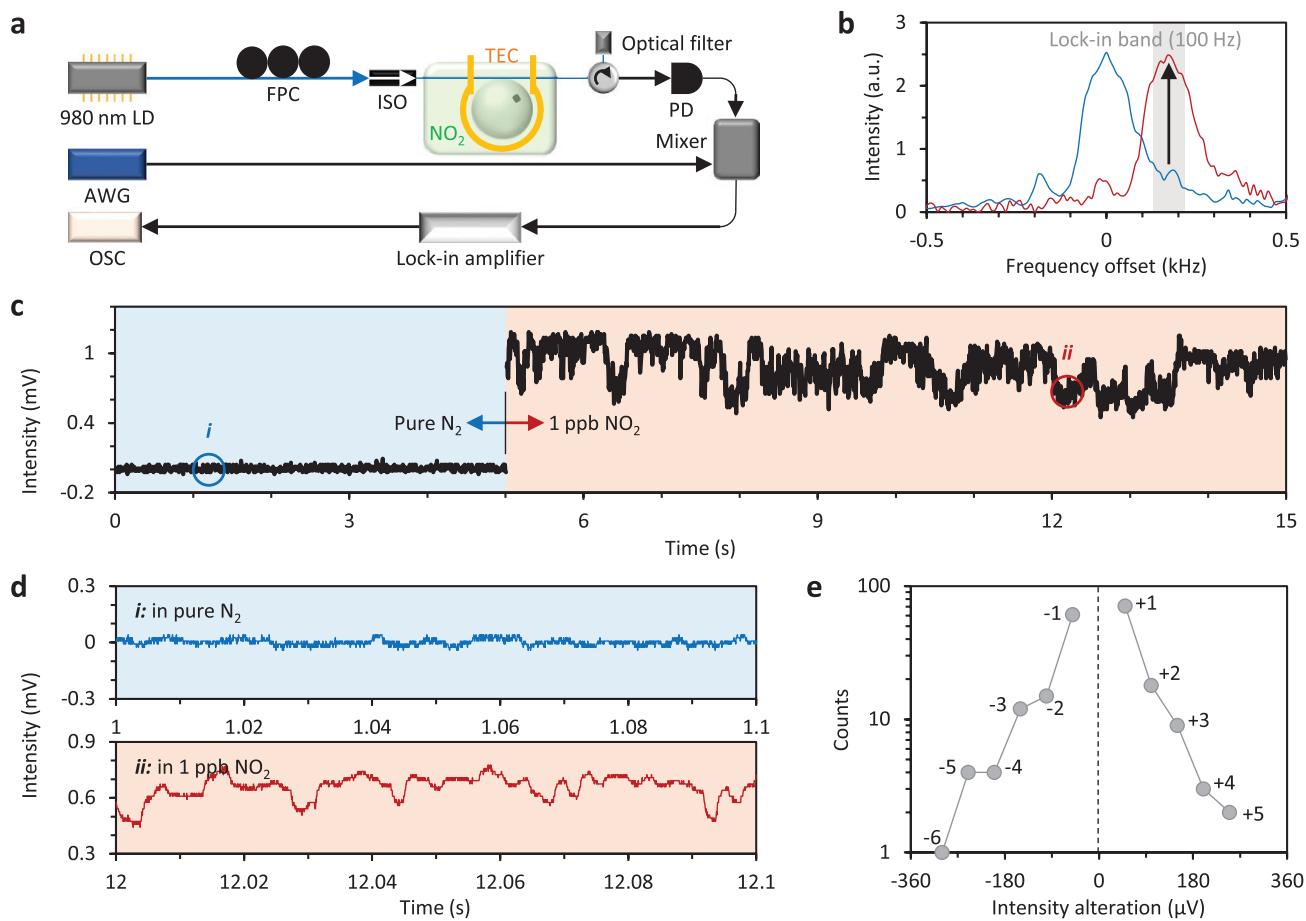

**Figure 5.** Measurement of individual molecule dynamics. a) Optoelectronic setup and lock-in detection for the measurement of individual molecule dynamics. b) The $f_{mix}$ signal shifts to the right when injecting 0.1 ppb of $NO_2$ into the chamber. Such spectral shift causes an intensity increment at the lock-in frequency. c) Temporal trace of the lock-in output. Here state i marks the response in pure $N_2$, while state ii marks the response in 1 ppb $NO_2$. d) Zoomed in traces from (c). Here the blue trace and the red trace show the device response in pure $N_2$ and in 1 ppb $NO_2$, respectively. A single molecule adsorption/desorption induces an intensity change of ±50 μV. e) Power-law statistics of the jumps observed in (d), indicating real-time measurement of individual gas molecule adsorption and desorption events.

*Gas Preparation*: Pure gas samples in sealed gas bags and standard reagent bottles were purchased. First, the gases were extracted by using quantitative syringes (Max range: 1 mL, ruler 0.1 μL). By repeatedly diluting and mixing the gas samples in dry $N_2$, the minimum injecting gas (0.1 μL, 1 ppm) could be controlled. In the experiment, the volume of the gas chamber with sealed fiber in–out channels was 8 L. The diluted gases were injected through a port connected with the interior of the closed gas chamber. The original filling gas in the chamber was pure $N_2$ at room temperature under the standard atmospheric pressure. Since the injected gas volume was much smaller than the gas chamber volume, the gas pressure change was negligible. In Note S3, Supporting Information, more details about the gas sensing measurements are shown.

## Supporting Information

Supporting Information is available from the Wiley Online Library or from the author.

## Acknowledgements

The authors acknowledge technical supports from Prof. Yuan Liu, Hunan University, and funding from the National Science Foundation of China (61975025 and U2130106) and the National Key Research and Development Program (2021YFB2800602). This work was also supported by the European Union's Horizon 2020 research and innovation program under Grant Agreement GrapheneCore3 881603. G.S. acknowledges the German Research Foundation DFG (CRC 1375 NOA project B5) and the Daimler und Benz foundation for financial support.

Open access funding enabled and organized by Projekt DEAL.

## Conflict of Interest

The authors declare no conflict of interest.

## Author Contributions

Ya.G., Z.L., and N.A. contributed equally to this work. B.Y. and Y.R. led this research. B.Y. and G.S. led the material organization. Ya.G. led the experimental measurements. Ya.G. and Z.L. performed the gas-sensing experiments. H.Z., N.A., and T.T. helped with the microsphere nanofabrication. Ya.G., Yo.G., and B.P. led the graphene exfoliation and transfer. C.W. and Y.W. contributed to the optoelectronic measurements. B.Y., Y.Y., and G.S. performed the theoretical analysis. All authors processed and analyzed the results. B.Y., Ya.G., G.S., and Y.R. prepared the manuscript with inputs from all co-authors.

## Data Availability Statement

The data that support the plots within this paper and other findings of this study are available from the corresponding author upon request.

## Keywords

graphene, microlaser sensors, mode splitting, multispecies gas detection, whispering gallery mode